\documentclass[twocolumn,prl]{revtex4-1}
\usepackage[latin9]{luainputenc}
\usepackage{color}
\usepackage{amsmath}
\usepackage{amssymb}
\usepackage{graphicx}
\usepackage[unicode=true,pdfusetitle,
 bookmarks=true,bookmarksnumbered=false,bookmarksopen=false,
 breaklinks=false,pdfborder={0 0 0},pdfborderstyle={},backref=false,colorlinks=true]
 {hyperref}
\hypersetup{
 colorlinks,linkcolor=red,citecolor=blue}

\makeatletter

\usepackage{amsfonts}

\makeatother

\begin{document}
\title{Non-reciprocal frequency conversion and mode routing in a microresonator}
\author{Zhen Shen$^{1,2}$}
\thanks{These authors contributed equally to this work.}
\author{Yan-Lei Zhang$^{1,2}$}
\thanks{These authors contributed equally to this work.}
\author{Yuan Chen$^{1,2}$}
\author{Yun-Feng Xiao$^{3}$}
\author{Chang-Ling Zou$^{1,2}$}
\author{Guang-Can Guo$^{1,2}$}
\author{Chun-Hua Dong$^{1,2}$}
\email{chunhua@ustc.edu.cn}
\affiliation{$^{1}$CAS Key Laboratory of Quantum Information, University of Science
and Technology of China, Hefei 230026, P. R. China.}
\affiliation{$^{2}$CAS Center For Excellence in Quantum Information and Quantum
Physics, University of Science and Technology of China, Hefei, Anhui
230026, P. R. China.}
\affiliation{$^{3}$State Key Laboratory for Mesoscopic Physics and Frontiers Science Center for Nano-optoelectronics, School of Physics, Peking University, Beijing 100871, China}

\date{\today}
\begin{abstract}
\textbf{The transportation of photons and phonons typically obeys the
principle of reciprocity. Breaking reciprocity of these bosonic excitations
will enable the corresponding non-reciprocal devices, such as isolators and circulators \cite{yu2009complete,shoji2014magneto,sounas2017non}. Here, we use two optical modes and
two mechanical modes in a microresonator to form a four-mode plaquette
via radiation pressure force. The phase-controlled non-reciprocal
routing between any two modes with completely different frequencies
is demonstrated, including the routing of phonon to phonon (MHz to
MHz), photon to phonon (THz to MHz), and especially photon to photon with frequency
difference of around 80 THz for the first time. In addition, one more mechanical mode is introduced to this plaquette to realize a phononic circulator in such single microresonator. The non-reciprocity is derived from interference between multi-mode transfer processes involving optomechanical interactions in an optomechanical resonator. It not only demonstrates the non-reciprocal routing of photons and phonons in a single resonator but also realizes the non-reciprocal frequency conversion for photons and circulation for phonons, laying a foundation for studying directional routing and thermal management in an optomechanical hybrid network.}
\end{abstract}
\maketitle
Optical and acoustic non-reciprocal devices are the basic building
blocks for information processing and sensing based on photons and
phonons. For
bosons, several methods can be used to achieve non-reciprocity,
ranging from applying a magnetic field \cite{bi2011chip,shoji2014magneto},
imposing rotational motion \cite{fleury2014sound,maayani2018flying,huang2018nonreciprocal,fu2019phononic},
optomechanical interactions \cite{verhagen2017optomechanical,xu2019nonreciprocal,mathew2020synthetic,chen2021synthetic,merklein2020chip}, spatiotemporal modulations \cite{lira2012electrically,tzuang2014non,dong2015brillouin,kim2015non,devaux2015asymmetric,sohn2018time,kittlaus2021electrically}, and chiral interaction of atoms \cite{scheucher2016quantum,zhang2018thermal,hu2020noiseless}. Magnetically induced non-reciprocity, as the most common mechanism, is
challenging to integrate low-optical-loss magnetic materials on chip for optical devices \cite{bi2011chip} and typically
produce mm-scale phononic devices because of weak non-reciprocal absorption \cite{lewis1972acoustic,sasaki2017nonreciprocal}.
The optomechanical
systems \cite{kippenberg2007cavity,aspelmeyer2014cavity}, are one of the most promising candidates to realize magnetic-free
non-reciprocity \cite{hafezi2012optomechanically,li2017optical},
where the optical isolation and non-reciprocal control of phonons have
been demonstrated respectively. Among them, one non-reciprocal mechanism is the use of phase matching conditions of traveling wave modes \cite{hafezi2012optomechanically,shen2016experimental,ruesink2016nonreciprocity,shen2018reconfigurable,kittlaus2018non}. Another mechanism is based on the gauge phase in the network of multiple nodes, which highlights the phase control of non-reciprocity \cite{fang2017generalized,ruesink2018optical,seif2018thermal,chen2021synthetic,xu2019nonreciprocal,mathew2020synthetic}.
Similar mechanisms have been applied to microwave photons in superconducting
circuits \cite{peterson2017demonstration,barzanjeh2017mechanical,bernier2017nonreciprocal,metelmann2017nonreciprocal,malz2018quantum,mercier2020nonreciprocal}.
Although the non-reciprocal mode conversion in superconducting circuits is demonstrated over a few gigahertz, applying the gauge phase mechanism to frequency dimension of optical photons and more modes in optomechanical systems will provide more non-trivial devices, such as phase-controlled non-reciprocal frequency conversion and phononic circulator.

\begin{figure}
\includegraphics[clip,width=1\columnwidth]{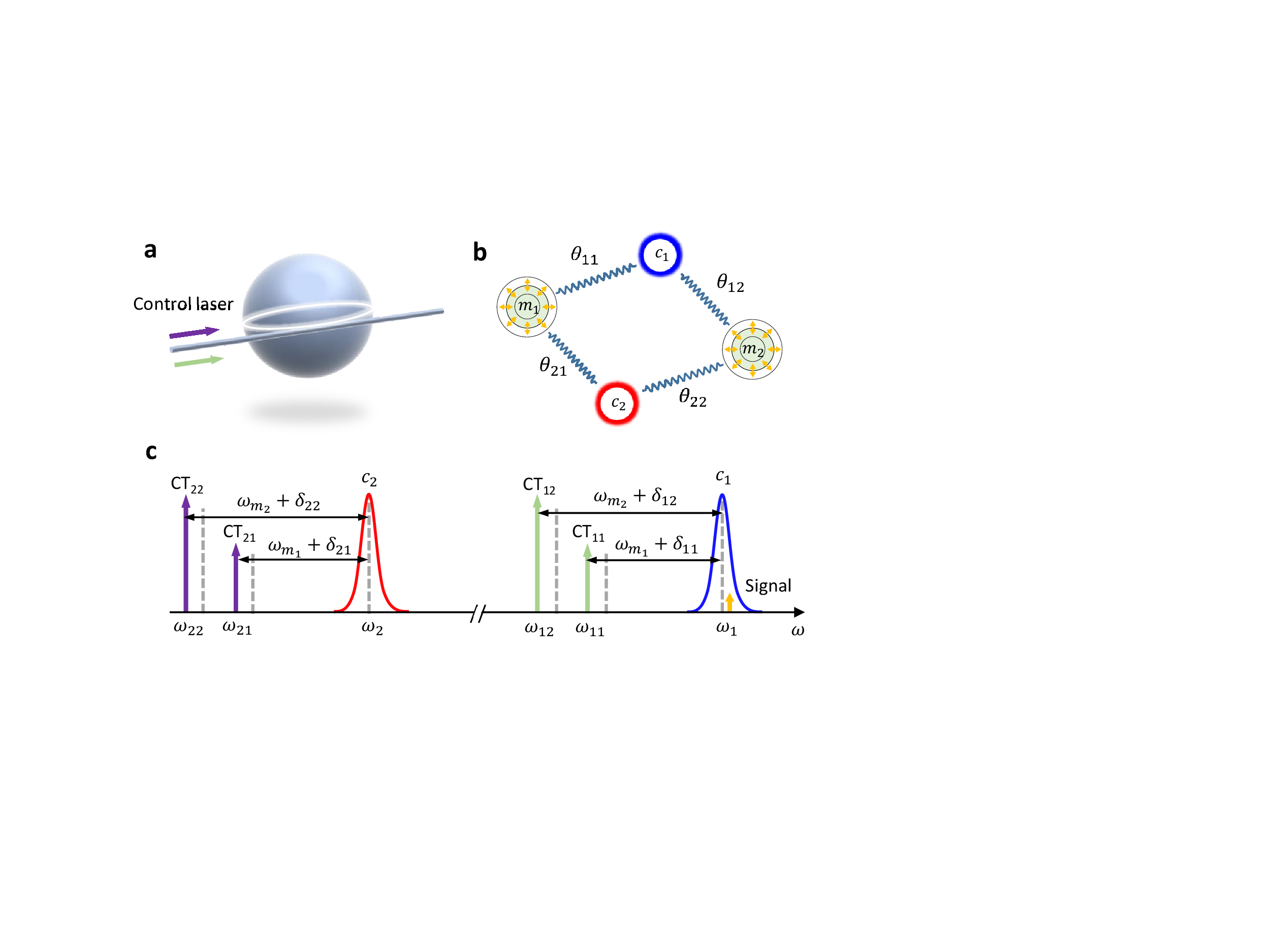}\caption{\textbf{Schematic of the non-reciprocal routing in a microresonator.} \textbf{a.} The control lasers enhance the optomechanical
coupling and control the non-reciprocal transportation inside a microsphere.
\textbf{b-c.} The typical two optical modes with frequencies $\omega_{1}$
and $\omega_{2}$ and two mechanical modes with frequencies $\omega_{\mathrm{m_{1}}}$
and $\omega_{\mathrm{m_{2}}}$ are used to establish a four-mode plaquette
via optomechanical interactions, which is controlled by four red-detuned
control fields. The non-reciprocal effects can be changed by tuning
the control phase $\theta_{11}$, $\theta_{12}$, $\theta_{21}$ and
$\theta_{22}$.}
\end{figure}

Here, we study the non-reciprocal routing of both photons and
phonons in a microresonator, where two optical modes and two mechanical
modes are exploited to form a closed loop of a four-mode plaquette,
using the optomechanical interactions as shown in Figs. 1a-b. The
four modes have completely different frequencies, i.e., 388 THz,
309 THz, 117 MHz, and 79 MHz, respectively. The non-reciprocal routing
between any two nodes among these four modes is demonstrated, including
the routing of phonon-phonon (MHz - MHz), photon-photon (THz - THz), and photon-phonon (THz - MHz). In general, the non-reciprocity results from the interference between two transportation paths engineered to connect the targeted nodes. The interference phase is governed by the phase of the control field, leading to a phase-controlled and flexible non-reciprocal routing. We demonstrate the phononic circulator simultaneously controlled by two independent phases when one more mechanical mode is introduced to such plaquette. The non-reciprocal frequency conversion opens new possibilities in communication and information processing, especially in the optical domain for wavelength-division-multiplexing (WDM) networks and quantum interface connecting separated systems operating at incompatible frequencies \cite{dong2012optomechanical,hill2012coherent,liu2013electromagnetically}. The phononic circulator holds great potential in thermal management, heat transfer, and phononics-based information processing \cite{xu2019nonreciprocal}.

\subsection*{Theoretical model}

The hybrid plaquette includes two optical modes and two mechanical
modes in the microresonator, as shown in Figs. 1b-c. The bosonic operators
$c_{\mathrm{1}}$ and $c_{\mathrm{2}}$ denote two optical whispering-gallery
modes with resonance frequencies of $\omega_{1}$, $\omega_{2}$ and
dissipation rates of $\kappa_{1}$, $\kappa_{2}$, respectively. The
bosonic operators $m_{1}$ and $m_{2}$ denote two mechanical modes
with resonance frequencies of $\omega_{\mathrm{m_{1}}}$, $\omega_{\mathrm{m_{2}}}$
and dissipation rates of $\gamma_{1}$, $\gamma_{2}$. We focus on achieving non-reciprocal routing between any two nodes of these four modes, that is, the bosonic excitations in one mode can be unidirectionally converted to another mode. A total of six pairs transportation ($m_{1}$-$m_{2}$, $c_{\mathrm{1}}$-$c_{\mathrm{2}}$,
$c_{\mathrm{\mathit{j}}}$-$m_{k}$ for $j,\:k\in\left\{ 1,2\right\} $)
should be respectively considered. Due to the symmetry,
we select three pairs of nodes ($m_{1}$-$m_{2}$, $c_{\mathrm{1}}$-$c_{\mathrm{2}}$,
$c_{\mathrm{1}}$-$m_{\mathrm{1}}$) for experimental verification,
showing that the non-reciprocal routing can be implemented between
phonon-phonon, photon-photon, and photon-phonon in the plaquette, where the non-reciprocal mechanism can be fully described by the general
theory of symmetry breaking between the two optomechanical coupling
paths. 

Here, we only theoretically describe the non-reciprocal routing of $m_{1}$-$m_{2}$ in detail (other non-reciprocal routings are discussed in the Supplemental
Material). When the cavity mode $c_{\mathrm{1}}$ is driven by two optical tones with frequencies of $\omega_{11}$ and $\omega_{12}$, the phonon of one mechanical mode up-converts to the optical mode $c_{\mathrm{1}}$ and then down-converts to the other mechanical mode. As this process of frequency conversion features reciprocity, the cavity mode $c_{\mathrm{2}}$ is introduced by the second pair of tones with frequencies of $\omega_{21}$ and $\omega_{22}$, to form another path of optically mediated frequency conversion simultaneously. Based on the path interference, the hopping phases controlled through external drives in such a plaquette give rise to the non-reciprocal routing \cite{peterson2017demonstration,barzanjeh2017mechanical,bernier2017nonreciprocal}.
In the frame rotating at the frequencies $\omega_{jk}$, the linearized Hamiltonian
can be written as ($\hbar=1$)
\begin{align}
H= & \sum_{j,k=1,2}G_{jk}\left[c_{j}^{\dagger}m_{k}e^{i\left(\delta_{jk}t+\theta_{jk}\right)}+c_{\mathrm{\mathit{j}}}m_{k}^{\dagger}e^{-i\left(\delta_{jk}t+\theta_{jk}\right)}\right],
\end{align}
where the driving fields are around the red sideband
of the cavity with the detuning $\delta_{jk}=\omega_{j}-\omega_{jk}-\omega_{m_{k}}$. And the counter-rotating terms can be neglected in the weak coupling
regime $G_{jk}\ll\omega_{m_{k}},\left|\omega_{m_{2}}-\omega_{m_{1}}\right|$.
$G_{jk}=g_{jk}\sqrt{n_{jk}}$ is the effective optomechanical interaction strength
with the vacuum optomechanical coupling rate $g_{jk}$ and the total
number of intracavity photons $n_{jk}$, and $\theta_{jk}$ is the
relative phase set by driving fields.

\begin{figure}
	\includegraphics[clip,width=1\columnwidth]{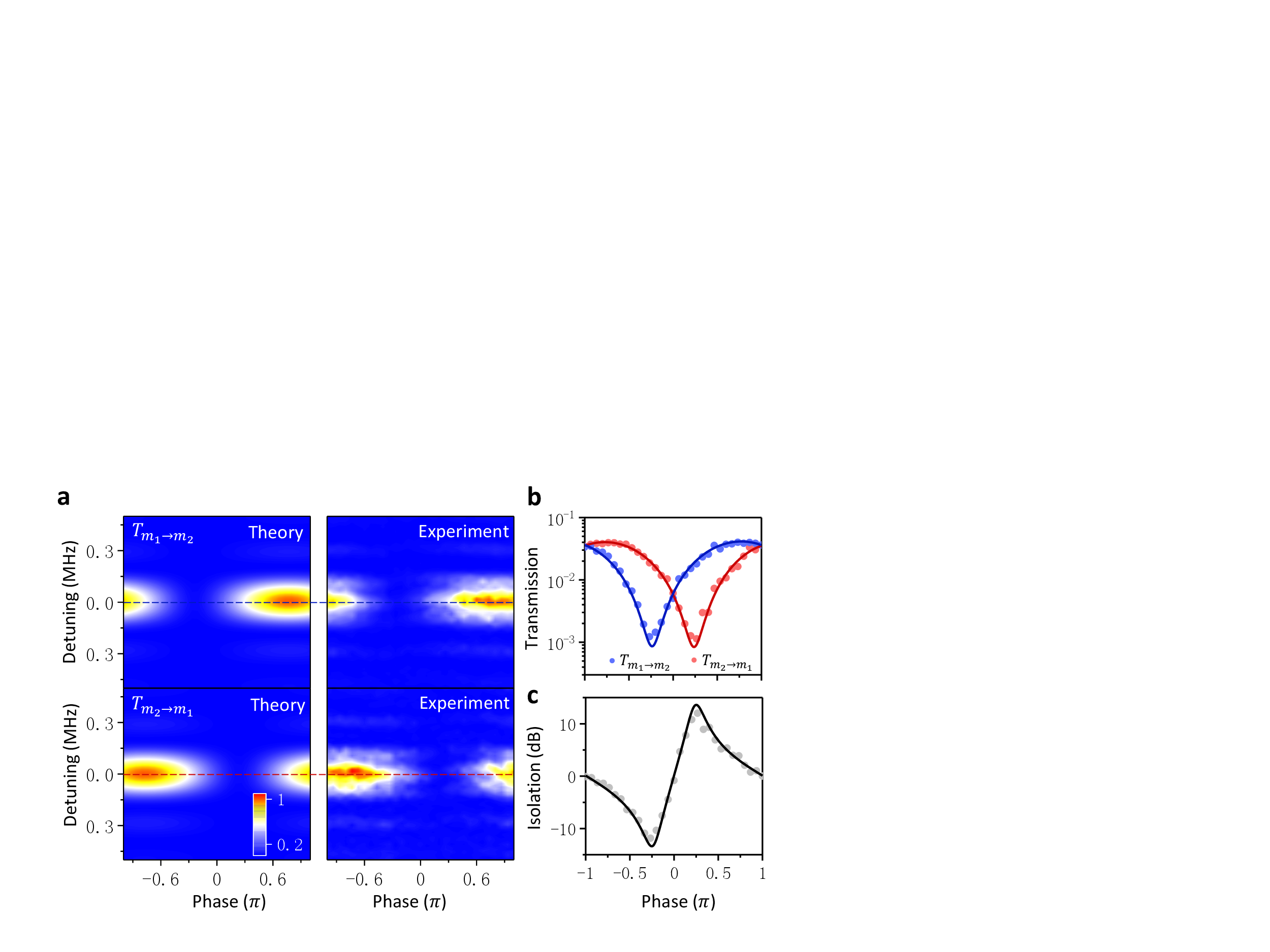}\caption{\textbf{Non-reciprocal routing between the two mechanical modes.}
		\textbf{a.} Theoretical and experimental normalized power spectra
		$m_{2,out}$ ($m_{1,out}$) converted from $m_{1,in}$ ($m_{2,in}$)
		as the function of controlled phase $\theta$. \textbf{b.} The transmissions
		$T_{m_{1}\rightarrow m_{2}}$ and $T_{m_{2}\rightarrow m_{1}}$ versus
		the control phase are obtained using the spectra data in \textbf{a}
		(the dashed blue line for $T_{m_{1}\rightarrow m_{2}}$ and the dashed
		red line for $T_{m_{2}\rightarrow m_{1}}$). \textbf{c.} The isolation
		$I=10log(T_{m_{1}\rightarrow m_{2}}/T_{m_{2}\rightarrow m_{1}})$
		as a function of the controlled phase, indicating more than 10 dB
		isolation in both directions in a reconfigurable manner. The solid
		lines in (b, c) indicate theoretically expected values.}
\end{figure}

For the interference of the two conversion paths, the effective phase $\delta_{jk}t+\theta_{jk}$ determines the non-reciprocity and conversion efficiency.
For the non-reciprocal transportation between two mechanical modes, we set the cavity detunings
$\delta_{11}=\delta_{12}=-\delta$ and $\delta_{21}=\delta_{22}=\delta$. And we define the transmission $T_{a\rightarrow b}=\left|b_{out}/a_{in}\right|^{2}$
with $a,~b\in\{m_{1},m_{2},c_{1},c_{2}\}$.
The ratio $\eta$ (See the Supplemental Material) of backward to forward transmission
reads

\begin{equation}
\eta=\frac{T_{m_{1}\rightarrow m_{2}}}{T_{m_{2}\rightarrow m_{1}}}=\left|\frac{G_{11}G_{12}e^{-i\theta}\chi_{1}\left(\omega\right)+G_{21}G_{22}\chi_{2}\left(\omega\right)}{G_{11}G_{12}e^{i\theta}\chi_{1}\left(\omega\right)+G_{21}G_{22}\chi_{2}\left(\omega\right)}\right|^{2},
\end{equation}
where $\theta=\theta_{11}+\theta_{12}-\theta_{21}-\theta_{22}$. $\omega$ is the detuning between the signal and mode resonance,  and $\chi_{j}^{-1}=\kappa_{j}/2-i\left[\omega+\left(-1\right)^{j}\delta\right]$
is the optical susceptibility with $j\in\{1,2\}$.
Obviously, we have the ratio $\eta=1$ when
$\theta=n\pi$ with $n=0,\pm1,\pm2,...$, which means that there is
no non-reciprocity for any detuning $\delta$ and frequency $\omega$.
When the cooperativities for all four optomechanical couplings are equal ($C=C_{jk}=4G_{jk}^{2}/\kappa_{j}\gamma_{k}$),
we can obtain the perfect non-reciprocity, i.e. $\eta=0$ or $\eta=\infty$, where the phase satisfies
\begin{equation}
\tan\left(\theta\right)=\pm\frac{\delta\left(\kappa_{1}+\kappa_{2}\right)+\omega\left(\kappa_{1}-\kappa_{2}\right)}{\kappa_{1}\kappa_{2}/2-2\left(\delta^{2}-\omega^{2}\right)}.
\end{equation}
In a practical experimental system, $\kappa_{1} \neq \kappa_{2}$ and the non-reciprocity still exists when $\delta=0$. However, we usually choose an appropriate $\delta$ in the experiment for an opitimized non-reciprocity. In addition, due to the symmetry of four-mode plaquette, the cavity detunings should be changed as $\delta_{11}=\delta_{21}=-\delta$ and $\delta_{12}=\delta_{22}=\delta$ for the photon-photon non-reciprocity.
For the adjacent conversion in such plaquette, i.e., phonon-photon conversion, the cavity detunings as mentioned before are all satisfied for the non-reciprocal routing (see Supplemental Material for more details).


\subsection*{Experimental realization}

To experimentally demonstrate the non-reciprocal routing between any two nodes in the four-modes plaquette, a silica microsphere
with a diameter of approximately $40\:\mu\textrm{m}$ is used in our experiment, where we choose
two whispering-galley modes with resonance frequency $\omega_{1}/2\pi=387.56\:\textrm{THz}$ (near $774\,\mathrm{nm}$), damping rates $\kappa_{1}/2\pi=7\:\textrm{MHz}$ and $\omega_{2}/2\pi=308.93\:\textrm{THz}$ ( near
$971\,\mathrm{nm}$), $\kappa_{2}/2\pi=27\:\textrm{MHz}$, respectively. The two radial breathing mechanical
modes have frequencies of $\omega_{\mathrm{m_{1}}}/2\pi=79\,\textrm{MHz}$
and $\omega_{\mathrm{m_{2}}}/2\pi=117\,\textrm{MHz}$ with dissipation
rates of $\gamma_{1}/2\pi=9\:\textrm{kHz}$ and $\gamma_{2}/2\pi=28\:\textrm{kHz}$,
respectively (see Supplementary Material for more details regarding
the setup).

Figure 2 shows the demonstrated experiment of non-reciprocal routing
between the two mechanical modes, where the resonator is driven with
four control fields ($CT_{11}$, $CT_{12}$, $CT_{21}$, $CT_{22}$) with powers ($P_{11}$, $P_{12}$, $P_{21}$, $P_{22}$) = (1.8, 3.5, 1.1, 4.5) mW that correspond to the cooperativities ($C_{11}$, $C_{12}$, $C_{21}$, $C_{22}$) = (3.6, 0.46, 0.73, 1).
The control fields are locked at a frequency detuning from the lower
motional sidebands of $\delta_{11}/2\pi=\delta_{12}/2\pi=-3\:\textrm{MHz}$
and $\delta_{21}/2\pi=\delta_{22}/2\pi=3\:\textrm{MHz}$.  Here, the
control lasers are respectively modulated by the acoustic-optic modulators
(AOMs) for pulse sequences to avoid thermal instability in the high-Q microresonator.
And the four control tones are all phase-locked through the RF
drives of AOMs, which can be varied continuously from $-\pi$
to $\pi$. As a first step, the phonons $m_{1,in}$(or $m_{2,in}$)
are prepared from the converted optical signal, which is resonant on the mode $c_{1}$ and converted by a red-detuned control tone $CT_{11}$ (or $CT_{12}$) served as writing pulse \cite{dong2012optomechanical, fiore2011storing,fiore2013optomechanical}. Then all control
tones are turned on simultaneously for an interaction duration $\tau=5\:\mu\textrm{s}$ to transfer the excitations from mode $m_{1}$ ($m_{2}$) to $m_{2}$ ($m_{1}$).
Finally, a control pulse $CT_{12}$ ($CT_{11}$) served as a readout pulse interacts with the converted mechanical excitations, converting the mechanical excitations back to an optical pulse at the cavity resonance, which corresponds to the retrieval process of the optomechanical light storage \cite{fiore2011storing,fiore2013optomechanical}.
Figure 2a shows the normalized
power spectra of the mode $m_{2,out}$(or $m_{1,out}$) transferred
from $m_{1,in}$(or $m_{2,in}$). The non-overlap transmittance unambiguously
present phase-controlled non-reciprocal routing: with $\theta=0.27\pi$
the excitations prepared in mode $m_{1}$ is transferred to mode $m_{2}$
($T_{m_{1}\rightarrow m_{2}}\approx1.8\%$) but not vice versa, while
with $\theta=-0.27\pi$ the excitations prepared in mode $m_{2}$
is transferred to mode $m_{1}$ but not vice versa. The average phonon occupation $N_{m_{2,out}}$ (or $N_{m_{1,out}}$) is estimated through the spectrally integrated area of the displacement power density spectrum of the mechanical excitation. We have used the thermal displacement power density spectrum and the corresponding average thermal phonon number, $N_{m_{2,th}}$ (or $N_{m_{1,th}}$), to calibrate the measurement.
The transmissions
$T_{m_{1}\rightarrow m_{2}}$ and $T_{m_{2}\rightarrow m_{1}}$ are
summarized and plotted in Fig. 2b as a function of the control phase
$\theta$ (see Supplementary Material for more details with regard
to the measurement). Therefore, isolation of more than 10 dB is demonstrated
in each direction in a reconfigurable manner, that is, the direction
of isolation can be switched by taking $\theta\rightarrow-\theta$,
as shown in Fig. 2c.
\begin{figure}
\includegraphics[clip,width=1\columnwidth]{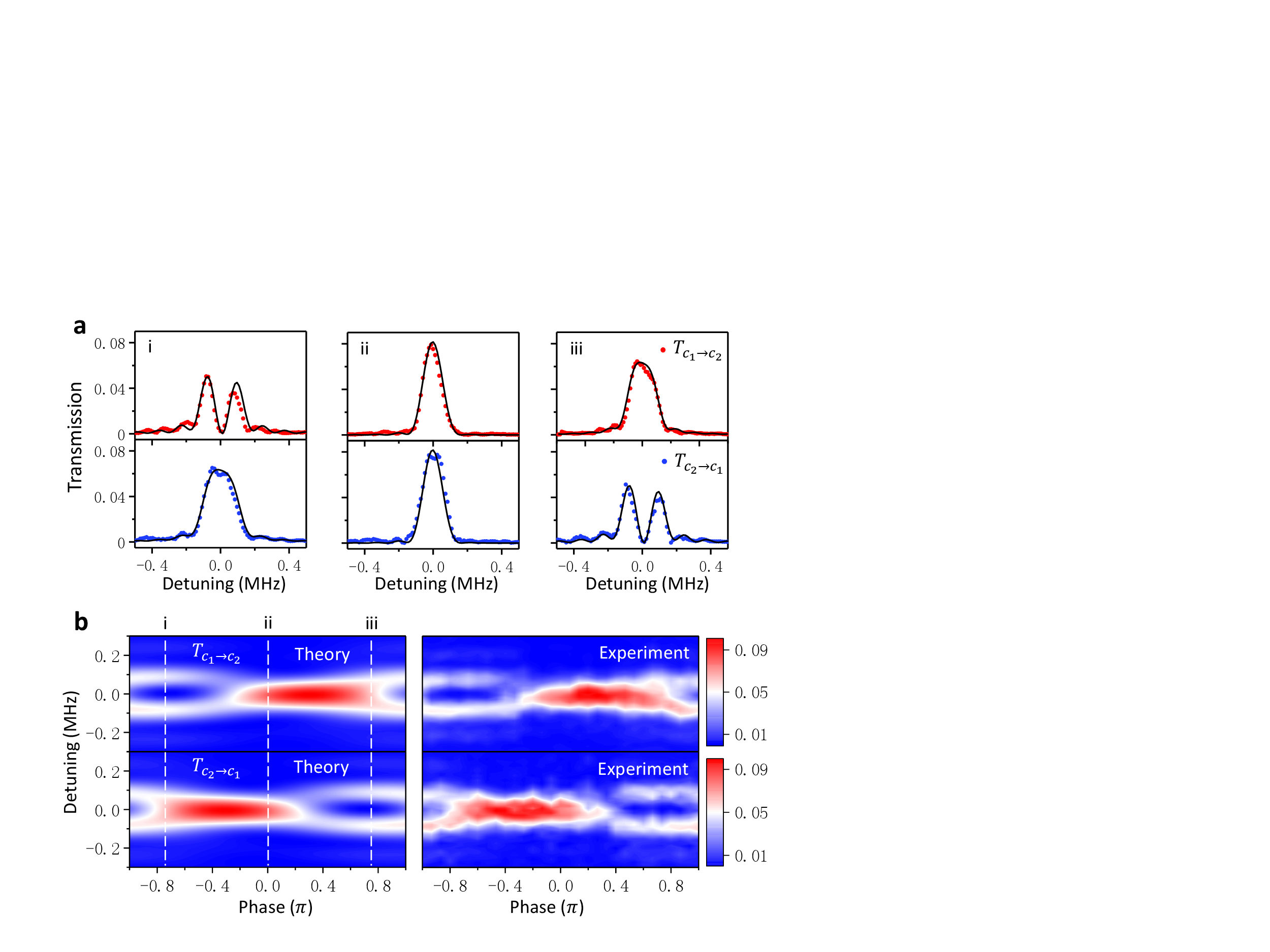}

\caption{\textbf{Non-reciprocal routing between the two optical modes}.
\textbf{a.} Measured transmission spectra $T_{c_{1}\rightarrow c_{2}}$
(red dots) and $T_{c_{2}\rightarrow c_{1}}$(blue dots) as a function
of probe detuning, shown in both directions for control phases $\theta=0.73\pi$,
0, $-0.73\pi$. The solid lines show the results of theoretical calculations.
Isolation of more than 15 dB in both directions is demonstrated with a bandwidth of 17.7 kHz, as
well as reciprocal behaviour with $\theta=0$. \textbf{b.} Theoretical
and experimental transmission spectra $T_{c_{1}\rightarrow c_{2}}$
and $T_{c_{2}\rightarrow c_{1}}$ as a function of the control phase and
probe detuning. Dashed lines correspond to the three measured phases
as shown in \textbf{a}.}
\end{figure}

Next, we experimentally realize the non-reciprocal routing
of photons between the two optical modes $c_{1}$ and $c_{2}$ through
two mechanical modes instead of the reciprocal photons conversion
via one mechanical mode \cite{dong2012optomechanical,shen2016experimental}.
At this point, the detunings of the four control tones are tuned to
a new configuration ($\delta_{\mathrm{11}}/2\pi=\delta_{\mathrm{21}}/2\pi=-45\:\textrm{kHz}$,
$\delta_{\mathrm{12}}/2\pi=\delta_{\mathrm{22}}/2\pi=45\:\textrm{kHz}$)
to establish the asymmetric paths achieving non-reciprocity. The pulse
sequence and the phase of the control field are generated similarly
with the case of non-reciprocal conversion of phonons (see Supplementary
Material for more details). The signal photons injected on resonance
with one optical mode will be converted to the other one, while four
control fields are synchronously to drive the mechanical motions. Frequency
conversion in both directions, $T_{c_{1}\rightarrow c_{2}}$ and $T_{c_{2}\rightarrow c_{1}}$
are measured and compared for three different phases in Fig. 3a. At
$\theta=0.73\pi$, we observe relatively high forward transmittance
(6\%) from cavity 1 to 2 and near-zero transmittance (0.1\%) in the
backward direction around zero detuning. Likewise, at the negative
phase of $\theta=-0.73\pi$ the transmission from cavity 1 to 2 is
suppressed while the transmission from cavity 2 to 1 is high. The peaks and dips with the linewidth of around $\sqrt{\gamma_{1}\gamma_{2}}$ observed in Fig. 2a, highlight the two-path interference mediated by two mechanical modes. Isolation of more than 15 dB in both directions is demonstrated with a bandwidth of 17.7 kHz. Around
$\theta=0$, the frequency conversion is reciprocal and bidirectional.
Figure 3b shows the transmission spectra as a function of probe detuning
for the whole range of phases $\theta$, where isolation of more than
15 dB is demonstrated in both directions. Here, we neglect the mode
coupling between clockwise (CW) and counter-clockwise (CCW) whispering-gallery
modes because of the weak backscattering. Actually, the traveling
wave cavity supports a pair of degenerate optical modes. The
CW modes $c_{1}$ and $c_{2}$ can be coupled with the CCW modes due
to the strong optical backscattering or nanoparticles \cite{zhu2010chip}.
Therefore, the model can be built including two paired optical modes. Finally, the non-reciprocal routing of bosonic excitations between the optical mode $c_{1}$ and the mechanical mode $m_{1}$ is demonstrated. The details of which are discussed in Supplementary Material. Due to the symmetry, the phase-controlled non-reciprocal routing between arbitrary two bosonic modes with ultra-high frequency difference
can be implemented.

The described four-mode plaquette can achieve flexible expansion, for example by parametrically coupling a third mechanical mode $m_{3}$ to the optical modes, as shown in Fig. 4a. This five-mode plaquette is established in another silica microsphere with $\omega_{\mathrm{m_{1}}}/2\pi=75.3\,\textrm{MHz}$, $\omega_{\mathrm{m_{2}}}/2\pi=76.4\,\textrm{MHz}$ and $\omega_{\mathrm{m_{3}}}/2\pi=112.7\,\textrm{MHz}$ (see Supplemental Material for more details). Similar to the four-mode case, we use six control fields with frequencies slightly detuned from the lower motional sidebands of the resonances to form the closed-loop. The powers of the control fields are ($P_{11}$, $P_{12}$, $P_{13}$, $P_{21}$, $P_{22}$, $P_{23}$) = (4.8, 1.6, 4.2, 3.4, 1.6, 1.7)mW corresponding to the cooperativities ($C_{11}$, $C_{12}$, $C_{13}$, $C_{21}$, $C_{22}$, $C_{23}$) = (1.6, 1.3, 1.3, 1.1, 0.86, 0.76). For an experimental demonstration of the phase-controlled phononic circulator, we measure the isolation $I_{ij}=10log(T_{m_{i}\rightarrow m_{j}}/T_{m_{j}\rightarrow m_{i}})$ ($i,$ $j\in\{1,2,3\}$) versus the control phases, as shown in Fig. 4b. Unlike the four-mode case, where the non-reciprocal conversion is achieved by varying only one control phase, in the phononic circulator, at least two independent control phases ($\theta_{11}$ and $\theta_{12}$ in our experiment) need to be tuned to change the phase difference of the three pairs of interference paths. As both $\theta_{11}$ and $\theta_{12}$ dominantly affect the transportation between mode $m_{1}$ and mode $m_{2}$, $I_{12}$ has a significant change in the diagonal direction, while $I_{23}$ ($I_{31}$) changes dominantly upon $\theta_{12}$ ($\theta_{11}$) varying, which is in agreement with intuition and theory (see Supplementary Material for more details). Figure 4c plots the $I_{sum}=I_{12}+I_{23}+I_{31}$, where we see phononic circulation in the forward direction $I_{sum}$ of up to 35 dB at ($\theta_{11}$, $\theta_{12}$) = (-0.7, 0.1)$\pi$ or in the backward direction $I_{sum}$ = -33.3 dB at ($\theta_{11}$, $\theta_{12}$) = (0.7, -0.1)$\pi$. Via a simple change in the control fields, the direction of circulator is reversed, demonstrating the great flexibility of our device.

Beyond the five-mode model, our demonstration can be scaled to the
two-dimensional hybrid network by exploiting more optical and mechanical
modes in the same microcavity, while each link couples one optical
mode and one mechanical mode. Indeed, the transmission of photons
between CW and CCW optical modes can also achieve non-reciprocal conversion
mediated by one mechanical mode \cite{chen2021synthetic}. Because
of the demonstration of non-reciprocal transportation between any
two nodes, we can control the propagation of the bosonic excitations
along the designated route (Supplementary Fig. S6). The expansion of the network in a single microcavity provides a potential platform for studying topological photonics/phononics and quantum many-body physics, without the requirement of fabricating massive identical microstructures.

In conclusion, we have experimentally demonstrated the non-reciprocal
routing of phonon-phonon, photon-photon, and photon-phonon simultaneously
in a  single microresonator. The non-reciprocal routing is controlled
by adjusting the frequency and phase of the control field. Among them,
the optical non-reciprocal frequency conversion can reach 6\% efficiency
and 15dB isolation, and more than 15dB isolation for the phonon non-reciprocal
conversion and circulation. These results can be applied to information processing
and thermal management, and also lay a foundation for the directional routing
of signals in a hybrid network.

\begin{figure}
\includegraphics[clip,width=1\columnwidth]{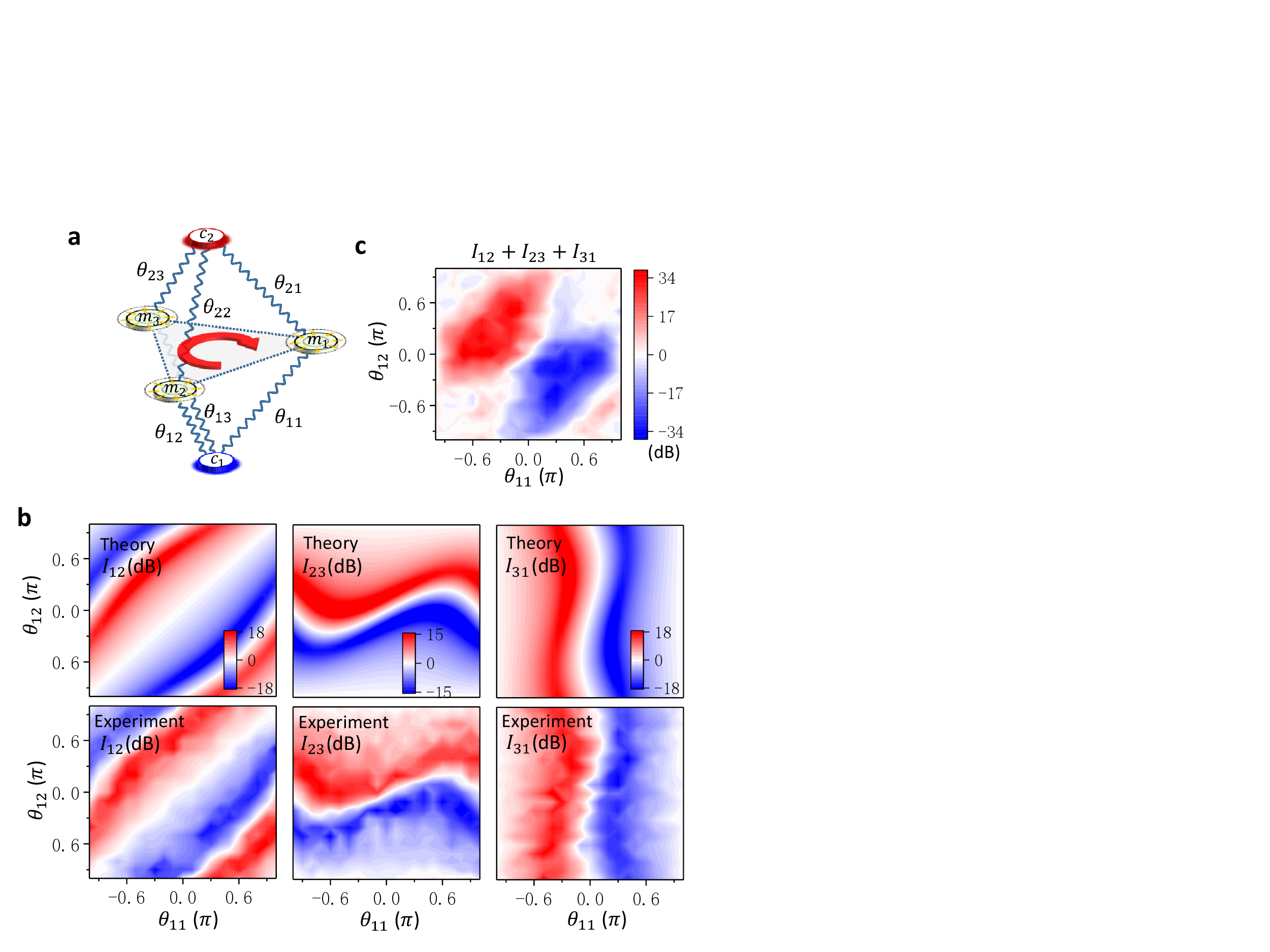}\caption{\textbf{Phononic circulator}.
\textbf{a}. Three mechanical modes $m_{1}$, $m_{2}$ and $m_{3}$ and two optical modes $c_{1}$ and $c_{2}$ are used to establish a five-mode network, creating a circulatory conversion between the three mechanical modes, as indicated by the red arrow. \textbf{b.} The measured isolation $I_{ij}=10log(T_{m_{i}\rightarrow m_{j}}/T_{m_{j}\rightarrow m_{i}})$ ($i,$ $j\in\{1,2,3\}$) and theoretical model as a function of control phases $\theta_{11}$ and $\theta_{12}$. \textbf{c.} The isolation $I_{12}+I_{23}+I_{31}$, indicating the circulator direction can be changed by control phases.}
\end{figure}

\vbox{}

\noindent\textbf{Acknowledgments}\\ The work was supported by the
National Key R\&D Program of China (Grant No.2016YFA0301303), the
National Natural Science Foundation of China (Grant Nos. 61805229, 11874342, 11934012, and 92050109), USTC Research Funds
of the Double First-Class Initiative (YD2470002002), the China Postdoctoral Science Foundation (Grant No. 2019M652181) and the Fundamental
Research Funds for the Central Universities. This work was partially
carried out at the USTC Center for Micro and Nanoscale Research and
Fabrication.

\vbox{}

\noindent\textbf{Author contributions}\\ Z.S. C.-L.Z. and C.-H.
D. conceived and designed the experiment. Z.S., Y.C. and C.H.D. prepared
the samples, built the experimental setup and carried out experiment
measurements. Y.-L.Z. Y.-F. X. and C.-L.Z. provided theoretical support and
analysed the data. Z.S., C.-H.D. and C.-L.Z. wrote the manuscript with inputs
from all authors. C.-H.D, and G.-C.G. supervised the project. 
All authors contributed extensively to the work presented in this
paper\emph{.}

\vbox{}

\noindent\textbf{Additional information}\\ Supplementary information
is available in the online version of the paper. Reprints and permissions
information is available online at. Correspondence and requests for
materials should be addressed to C.-H.D. (chunhua@ustc.edu.cn).

\vbox{}

\noindent\textbf{Competing financial interests}\\The authors declare
no competing interests.

\end{document}